\documentstyle[preprint,revtex]{aps}
\begin{document}
\draft
\preprint{IPT-EPFL November 1993}
\begin{title}
Invariants of the $1/r^2$ Supersymmetric t-J Models
\end{title}
\author{D. F. Wang and C. Gruber}
\begin{instit}
Institut de Physique Th\'eorique\\
\'Ecole Polytechnique F\'ed\'erale de Lausanne\\
PHB-Ecublens, CH-1015 Lausanne-Switzerland.
\end{instit}
\begin{abstract}
In this work, we have studied the invariants of motion of two $SU(N)$
supersymmetric t-J models of $1/r^2$ hopping and exchange
in one dimension.
The first model is defined on a lattice of equal spaced sites,
and the second on a non-equal spacing lattice.
Using the ``exchange operator formalism'',
we are able to construct all the invariants for the models, by
mapping the systems to mixtures of fermions and bosons.
This identification shows that the supersymmetric t-J model on the chain
with equal-spaced sites also belongs to
Shastry-Sutherland's ``Super-Lax-Pair'' family.
\end{abstract}

\narrowtext
Since the independent works
by Haldane and Shastry, there have been renewed interests in
exactly solvable models of long range
interaction\cite{hald88,shas88,hald91,shastry,shastry2,kura91,kawakami92,wang,fowler,poly,frahm,suth72,rucken92,ha,coleman,wang3}.
Of these systems, the 1D supersymmetric t-J model
of $1/r^2$ exchange and hopping
has been studied intensively\cite{kura91,kawakami92,wang,ha}.
The system is identified as
a free Luttinger liquid\cite{kura91,kawakami92,wang},
and the asymptotic correlation functions have been
calculated through finite size scaling technique.
The excitation spectrum of the system may be obtained
with the help of the asymptotic Bethe-ansatz.
In particular, for the $SU(2)$ case, the asymptotic Bethe-ansatz
spectrum was explicitly shown to be exact, and the correct
thermodynamics was given when the spinon rotation was properly
taken into account\cite{wang}. In general, exact solvability implies existence
of infinite number of constants of motion.
For the long range t-J models, the complete construction of invariants
of motion has remained unknown. In this work, applying the
``exchange operator
formalism'' to a mixture of fermions and bosons,
we are able to systematically provide all the invariants for the
$SU(N)$ systems.

Let us first consider the $1/r^2$ supersymmetric t-J model
on a one dimensional lattice of equal spaced sites.
The Hamiltonian for the one-dimensional t-J model is given by
\begin{equation}
H=P_{G}
\left[ -\sum_{1\le i\ne j \le L}\sum_{\sigma=1}^{N} t_{ij} ( c_{i\sigma}
^{\dagger}
c_{j\sigma}) +
\sum_{1\le i\ne j \le L}
J_{ij} \left[ P_{ij}-(1-n_i)(1-n_j)\right] \right] P_{G},
\label{eq:original}
\end{equation}
where we take the hopping matrix and the spin exchange interaction to be
$t_{ij}/2 = J_{ij} = 1/d^2(i-j)$, and $d(n)=(L/\pi) \sin (n\pi/L)$
is the chord distance, with $L$ the size of the lattice.
The operator $c_{i\sigma}^{\dagger}$ is the
fermionic operator to create an electron with spin component $\sigma$
at site $i$, $c_{i\sigma}$ is the corresponding
fermionic annihilation operator. Their anti-commutation relations
are given by $\{c_{i\sigma_i},c_{j\sigma_j}^{\dagger}\}_+
=\delta_{ij} \delta_{\sigma_i\sigma_j},
\{c_{i\sigma_i},c_{j\sigma_j}\}_+=0, \{c_{i\sigma_i}^\dagger
,c_{j\sigma_j}^\dagger\}_+=0$.
We assume that the spin component
$\sigma$ takes values from $1$ to $N$.
The Gutzwiller projection operator $P_{G}$
projects out all the double or multiple occupancies,
$P_{G} = \prod_{i=1}^L P_{G}(i)$, and $P_{G}(i) =\delta_{0,n_i}
+\delta_{1,n_i}$, with $n_i=\sum_{\sigma=1}^N c_{i\sigma}^\dagger
c_{i\sigma}$.
The operator $P_{ij}=\sum_{\sigma=1}^N \sum_{\sigma'=1}^N
c_{i\sigma}^{\dagger} c_{i\sigma'} c_{j\sigma'}^{\dagger} c_{j\sigma}$
exchanges the spins of the electrons at sites $i$ and $j$, if both sites
are occupied. $n_i$ and $n_j$ are the electron number operators
at sites $i$ and $j$.

Now, on the lattice, we may introduce two new fields, the $f$ and
$b$ fields.
For the new fields, we have $\{f_{i\sigma},f_{j\sigma'}\}_{+} =0,
\{f_{i\sigma}^{\dagger}, f_{j\sigma'} \}_{+} =
\delta_{ij} \delta_{\sigma\sigma'}, \left[ b_i, b_j \right]=0,
\left[ b_i, b_j^\dagger \right] =\delta_{ij} $. The $b$ field
always commute with the $f$ field. The size of the Hilbert space at
each site is $\infty$ in this case. However, let us project out
the zero occupancy and
all the double or multiple occupancy, and we work in the subspace
where there is only one particle at each site.
This new subspace can
be shown to be equivalent to the subspace
defined by the $c$ field with no double or multiple occupancies.
In particular, we may represent the fermionic electron operators
$c_{i\sigma}^{\dagger}$ and $c_{i\sigma}$ in the following way:
\begin{eqnarray}
&& P_{G}(i) c_{ i \sigma}^{\dagger} P_{G}(i) =
\delta_{1,n_b^i+n_f^i} f_{i \sigma}^{\dagger} b_i \delta_{1,n_b^i+n_f^i}
\nonumber\\
&& P_{G}(i) c_{i\sigma} P_{G}(i)
= \delta_{1,n_b^i+n_f^i} b_i^{\dagger} f_{i\sigma}
\delta_{1,n_b^i+n_f^i},
\label{eq:mapping}
\end{eqnarray}
where $n_b^i +n_f^i = b_i^{\dagger} b_i +
\sum_{\sigma =1}^{N} f_{i\sigma}^{\dagger} f_{i\sigma}$.
In terms of the
$f$ and $b$ fields, a state vector can be written as
\begin{eqnarray}
|\phi> = \sum_{\sigma_1, \sigma_2, \cdots, \sigma_{N_e} }
&&\sum_{ \{x\},\{y\} } \phi ( x_1\sigma_1, x_2\sigma_2,
\cdots, x_{N_e}\sigma_{N_e} | y_1, y_2, \cdots, y_Q )\times \nonumber\\
\times &&f_{x_1\sigma_1}^{\dagger} f_{x_2\sigma_2}^{\dagger} \cdots
f_{x_{N_e} \sigma_{N_e}}^{\dagger} b_{y_1}^{\dagger} b_{y_2}^{\dagger}
\cdots b_{y_Q}^{\dagger} |0>,
\label{eq:amp}
\end{eqnarray}
where $N_e$ is the number of $f$ fermions on the lattice, $Q$ is the
number of $b$ bosons, and we require that
$x_i\ne x_j \ne y_k \ne y_l$, and that
the $f$ fermion positions $\{x\}$ and the $b$
boson positions $\{y\}$ span the whole chain.
Obviously, $N_e$ is also the number of
the electrons, and $Q$ is also the number of holes on the lattice.
The amplitude $\phi$ is anti-symmetric when exchanging
$(x_i\sigma_i)$ and $ (x_j\sigma_j)$, and symmetric in the
boson coordinates $\{y\} = (y_1, y_2, \cdots, y_Q)$.
The Hamiltonian of the supersymmetric t-J model, using the
mapping Eq.~(\ref{eq:mapping}) in a straightforward way, can be
written in terms of
the fermionic $f$ field and the bosonic $b$ field.

With the above mapping, we may write the eigen-energy equation of the
supersymmetric t-J model in first quantized form.
Define the ``exchange operator'' $M_{ij}$ as
$M_{ij} F(q_1,q_2,\cdots, q_i, \cdots, q_j, \cdots, q_L )
= F(q_1,q_2, \cdots, q_j, \cdots, q_i, \cdots, q_L)$,
where the function $F$ is an arbitrary function of some position variables
$(q_1, q_2, \cdots, q_L)$, i.e., the operator $M_{ij}$
exchanges the positions $q_i, q_j$
of the particles $i $ and $j $.
In terms of such exchange operators, the eigen-energy equation of the
t-J model takes the form as follows
\begin{equation}
-\left[ \sum_{1\le i\ne j\le L} d^{-2} (q_i -q_j)
M_{ij} \right]
\phi(\{q\};\{\sigma\} ) = E \phi(\{q\};\{\sigma\}),
\label{eq:eigen}
\end{equation}
where $ \{ q \}=(q_1, q_2, \cdots, q_L) =
(x_1, x_2, \cdots, x_{N_e}, y_1, y_2, \cdots, y_Q)$,
and $\phi(\{q\};\{\sigma\})
= \phi(q_1\sigma_1,q_2\sigma_2,\cdots,q_{N_e}\sigma_{N_e}
|q_{N_e+1} q_{N_e+2} \cdots q_{L})=\phi (x_1\sigma_1, x_2\sigma_2,
\cdots, x_{N_e}\sigma_{N_e} | y_1, y_2, \cdots, y_Q)$ is the
amplitude of the state vector of Eq.~(\ref{eq:amp}).
$\{\sigma\} = (\sigma_1, \sigma_2, \cdots, \sigma_{N_e})$
are the spin variables of the $f$ fermions.
The operation $M_{ij}$ is defined in the conventional way:
$M_{ij} \phi(\{q\};\{\sigma\})
=\phi (\{q'\};\{\sigma\}) $, with $\{q\}=(q_1,q_2,\cdots,q_i,\cdots,q_j,
\cdots, q_L)$ and $\{q'\}=(q_1,q_2,\cdots,q_j,\cdots,q_i,\cdots,q_L)$.
Here, the sum in the Eq.~(\ref{eq:eigen}) is over all pairs of the particles.
Thus we see that using the $f$ and $b$ fields, we
may write the original t-J model as an eigen-value problem
for a mixture of the $f$ fermions and the spinless $b$ bosons
in terms of the ``exchange operators''.

Recently, Fowler and Minahan have considered a gas of identical
bosons on one dimensional chain\cite{fowler}.
Using the so-called ``exchange operator
formalism'', they have been able to explicitly construct all
the invariants of motion for the $SU(N)$ spin chains of Haldane-Shastry.
Let us briefly review their results.
Say, $M_{ij}$ is the exchange
operator that interchanges $q_i$ and $q_j$,
the positions of the particle $i$ and
particle $j$, when we operate $M_{ij}$ on a wavefunction
$F(q_1, q_2, \cdots, q_{L}) $.
In terms of this operator, they have been able to construct
an infinite sets of quantities $I_n$ that commute among themselves,
\begin{equation}
\left[ I_n, I_m \right] = 0,
\end{equation}
where $I_n = \sum_{i=1}^L \pi_i^n $,
with $\pi_i = \sum_{ j(\ne i)} (z_j/z_{ij}) M_{ij} $,
$z_i = e^{ 2\pi i q_i /L}, z_{ij} = z_i -z_j$,
and $n, m =0, 1, 2, \cdots, \infty$.
It was found that
all these quantities commute with the
Hamiltonian $H = \sum_{1\le i\ne j \le L} |z_i -z_j|^{-2} M_{ij}$
as long as the particles
occupy the whole chain. For a system of identical bosons on the chain,
the wavefunction is totally symmetric when we simultaneously interchange
spins and positions of two particles. The effect of the exchange
operator $M_{ij}$ is just equivalent to the effect of
the spin exchange operator alone. Using this method, they have
successfully constructed all the invariants for the $SU(N)$
Haldane-Shastry model.

We would like to stress that, in the language of the
exchange operators $M_{ij}$, the commutation
results proved by Fowler and Minahan
hold as operator identities. The central issue is that the forms
of wavefunctions of many particle systems, as well as the statistics of the
particles or the types of the particles, do $\it not $ matter in order
for the commutators to hold, as long as the particles occupy the whole
chain. We may then apply the ``exchange operator formalism''
to the wavefunctions of mixtures of fermions and bosons.
Therefore, from the eigen-equation Eq.~(\ref{eq:eigen}),
we thus conclude that in the first quantization,
all the invariants of the t-J model are the same
$I_n$'s as constructed by Fowler and Minahan,
which can be written in terms of the
exchange operators $M_{ij}$'s.

With the permutation properties of the amplitude $\phi$ for the mixture of
bosons and fermions, it is straightforward to
write all the invariants of motion of the t-J model in the
second quantization form using the $I_n$'s.
For instance, the exchange operation between the $f$ fermion positions
is equivalent to the spin exchange operation ( minus sign involved ),
the exchange operation between $b$ boson positions is equivalent to
the hole-hole interaction term, and exchange operation between
$f$ fermion and $b$ boson positions is equivalent to the
electron hopping. Such procedure to reduce an $I_n$ to a second
quantized form is quite simple, and
we do not write all the details.
Thus we provide a systematical way to construct all
the invariants of motion for the $1/r^2$
supersymmetric t-J model,
either in first quantized or in second quantized forms.

Recently, Shastry and Sutherland have studied the
interesting relation between supersymmetry and integrability, through
the so-called ``Super-Lax-Pair'' approach\cite{shastry,shastry2}.
For this equal-spacing chain, using the
mapping Eq.~(\ref{eq:mapping}), we have been able to write the
t-J model Hamiltonian Eq.~(\ref{eq:original})
in terms of the exchange operators
as Eq.~(\ref{eq:eigen}). This identification shows that the ``Super-Lax-Pair''
results obtained by Shastry and Sutherland
may apply to this t-J model\cite{shastry}.

Besides the above integrable t-J model on equally-spaced sites,
let us consider another supersymmetric
t-J model of $1/r^2$ hopping and exchange
on a chain with sites not equally spaced.
The positions of the sites $x_1, x_2, \cdots, x_{L}$ are determined
by the equation
\begin{equation}
x_i = \sum_{1\le j (\ne i)\le L} 2/(x_i-x_j)^3.
\label{eq:positions}
\end{equation}
This equation has appeared before in a paper discussing
a long range spin chain of Haldane-Shastry type. Doping
this spin chain, we are led to the following supersymmetric
t-J model
\begin{equation}
H=P_{G} \left[
-\sum_{1\le i\ne j \le L} \sum_{\sigma=1}^{N} t_{ij} ( c_{i\sigma}
^{\dagger}
c_{j\sigma}) +\sum_{1\le i\ne j \le L}
J_{ij} \left[ P_{ij}-(1-n_i)(1-n_j)\right] \right] P_{G},
\end{equation}
where the hopping matrix and the anti-ferromagnetic
exchange interaction are given by $t_{ij}/2 = J_{ij} = 1/(x_i-x_j)^2$, and
each site is occupied at most by one electron.

In the half filling case $N_e = L$, this system reduces to the spin chain
that has been studied before,
which is completely solvable and
similar exchange operator formalism has been developed\cite{poly,frahm}.
Let us just write down the results obtained by Polychronakos:
$[I_n, I_m] = 0, [I_n, H]=0$, where $I_n = \sum_{i=1}^L h_i^n$,
$h_i=a_i^{\dagger} a_i$, and
$a_i^{\dagger} = \pi_i^{\dagger} + i q_i, a_i=\pi_i -iq_i$, with
$\pi_i =\sum_{j(\ne i)} i(q_i-q_j)^{-1} M_{ij}, H=\sum_{i\ne j}
(q_i-q_j)^{-2} M_{ij} $, and $n, m=0,1,2,\cdots, \infty$.
Here, all the particles are put on the chain where the sites
are positioned as determined by the Eq.~(\ref{eq:positions}).
We may relate the operation of exchanging particle positions
to the operation to exchange particle spins, by assuming
that we have identical bosons again, for which the wavefunctions
are totally symmetric when we exchange the spins and positions of
two particles simultaneously. With this assumption,
from $I_n$'s, we thus can derive all the invariants of motion
written in terms of the spin exchange operators alone.

Again, all commutation results written in terms of the exchange operators
$M_{ij}$ obtained by Polychronakos
hold as operator identities, as long as the particles occupy the whole chain
of the sites positioned in the special way.
The forms of the wavefunctions
do ${\it not}$ matter at all. Thus the commutation results
may apply to wavefunctions of
particles of arbitrary statistics or wavefunctions of
mixtures of particles of different statistics on the chain.
Mapping our new supersymmetric t-J model
in terms of the $b$ and $f$ fields,
we may also write the eigen-energy equation in
first quantized form. In terms of the exchange operators between
the positions of the bosons and fermions, the Hamiltonian takes
the form
\begin{equation}
H=-\sum_{1\le i\ne j \le L} (q_i-q_j)^{-2} M_{ij}.
\label{eq:eigen2}
\end{equation}
Applying the formalism to this t-J model, in a similar way
we may obtain all the invariants, either in first quantized or
in the second quantized form, which commute among themselves and
with the Hamiltonian. Thus this supersymmetric t-J model is
also completely integrable.

In conclusion, we have studied the invariants of motion of two $SU(N)$
supersymmetric t-J models of long range hopping and exchange.
The first system is on the chain of equal spaced sites, and
the other on a chain of non-equal spaced sites.
Mapping the corresponding t-J model Hamiltonians
to those written in terms of mixed fermionic and bosonic
fields, then applying the ``exchange operator formalism'',
we for the first time are able to construct all the invariants
of the original Hamiltonian systematically.

Finally, we wish to point out that, for the second t-J model,
we will also have a metal-insulator phase transition at half-filling.
Away from half filling, we would also expect to have decoupled
spin and charge excitations near the ground state.
The system would be a Luttinger liquid.
The study of the physical properties
of the t-J model, such as
its full excitation spectrum and finite temperature properties,
is reported in our forthcoming paper.
We also have Jastrow product
ground state and excited state wavefunctions\cite{gruber}, as in the case
for the model on the chain with equally spaced sites. It would also
be very interesting to find out possible Shastry-Sutherland type
``Super-Lax-Pair'' for this new supersymmetric t-J model on
non-equal spacing chain.
We also hope to return to the issue of constructing invariants
of the non-supersymmetric t-J models of the $1/r^2$ hopping
and exchange in future.

This work was supported by the World Laboratory.


\begin{references}
\bibitem{hald88} F. D. M. Haldane, Phys. Rev. Lett. {\bf 60}, 635 (1988).
\bibitem{shas88} B. S. Shastry, Phys. Rev. Lett. {\bf 60}, 639 (1988),
Phys. Rev. Lett. {\bf 69}, 164 (1992).
\bibitem{hald91} F. D. M. Haldane, Phys. Rev. Lett. {\bf 66}, 1529 (1991).
\bibitem{shastry} B. S. Shastry and B. Sutherland,
Phys. Rev. Lett. {\bf 70}, 4092 (1993).
\bibitem{shastry2} B. Sutherland and B. S. Shastry,
Phys. Rev. Lett. {\bf 75}, 5 (1993).
\bibitem{kura91} Y. Kuramoto and H. Yokoyama, Phys. Rev. Lett. {\bf
67}, 1338 (1991).
\bibitem{kawakami92} N. Kawakami, Phys. Rev. B {\bf 45}, 7525 (1992).
\bibitem{wang} D. F. Wang, James T. Liu and P. Coleman,
Phys. Rev. B {\bf 46}, 4663 (1992).
\bibitem{fowler} M. Fowler and J. A. Minahan, Phys. Rev. Lett.
{\bf 70}, 2325 (1993).
\bibitem{poly} A. P. Polychronakos, Phys. Rev. Lett. {\bf 70},
2329 (1993).
\bibitem{frahm} Holger Frahm, J. Phys. A. {\bf 26}, 473 (1993).
\bibitem{suth72} B. Sutherland, Phys. Rev. A {\bf 5}, 1372 (1972),
Phys. Rev. A {\bf 4}, 2019 (1971), J. Math. Phys. {\bf 12}, 251 (1971),
J. Math. Phys. {\bf 12}, 246 (1971),
\bibitem{gruber} C. Gruber and D. F. Wang, IPT-EPFL preprint,
{\it ``Exact Results of the 1D $1/r^2$ Supersymmetric t-J Model
without Translational Invariance''}, November 1 (1993).
\bibitem{rucken92} F. Gebhard and A. E. Ruckenstein, Phys.
Rev. Lett. {\bf 68}, 244 (1992).
\bibitem{ha} Z. Ha and Haldane, Phys. Rev. {\bf B 46}, 9359 (1992).
\bibitem{coleman} D. F. Wang, Q. F. Zhong and P . Coleman,
Phys. Rev. {\bf B 48}, 8476 (1993).
\bibitem{wang3} D. F. Wang, Phys. Rev. {\bf B 48}, 10556 (1993).
\end{references}
\end{document}